\documentclass{webofc}
\usepackage[varg]{txfonts}   

\usepackage[T1]{fontenc}
\usepackage{pifont}  
\usepackage{tikz}
\usepackage[per-mode=symbol,bracket-unit-denominator=false,detect-all,load-configurations=binary,binary-units]{siunitx}
\DeclareSIUnit\loc{LOC}
\DeclareSIUnit\permil{\textperthousand}
\usepackage[frozencache]{minted}
\usetikzlibrary{arrows,positioning,shapes,topaths,calc,fit,backgrounds,matrix,shadows}

\begin{document}
\title{Evolution of the ROOT Tree I/O}
%
%

\author{\firstname{Jakob} \lastname{Blomer}\inst{1}\fnsep\thanks{\email{jblomer@cern.ch}} \and
        \firstname{Philippe} \lastname{Canal}\inst{2} \and
        \firstname{Axel} \lastname{Naumann}\inst{1} \and
        \firstname{Danilo} \lastname{Piparo}\inst{1}
}

\institute{CERN, Geneva, Switzerland
\and
           Fermilab, Chicago, U.S.
}

\abstract{%
     The ROOT TTree data format encodes hundreds of petabytes of High Energy and Nuclear Physics events.
     Its columnar layout drives rapid analyses, as only those parts (``branches'') that are really used in a given analysis need to be read from storage.
     Its unique feature is the seamless C++ integration, which allows users to directly store their event classes without explicitly defining data schemas.
     In this contribution, we present the status and plans of the future ROOT 7 event I/O.
     Along with the ROOT 7 interface modernization, we aim for robust, where possible compile-time safe C++ interfaces to read and write event data.
     On the performance side, we show first benchmarks using ROOT's new experimental I/O subsystem that combines the best of TTrees with recent advances in columnar data formats.
     A core ingredient is a strong separation of the high-level logical data layout (C++ classes) from the low-level physical data layout (storage backed nested vectors of simple types).
     We show how the new, optimized physical data layout speeds up serialization and deserialization and facilitates parallel, vectorized and bulk operations.
     This lets ROOT I/O run optimally on the upcoming ultra-fast NVRAM storage devices, as well as file-less storage systems such as object stores.
}
\maketitle

\section{Introduction}
\label{sct:intro}

The data describing a High Energy Physics (HEP) event is typically represented by a record containing variable-length collections of sub records.
An event can, for instance, contain a collection of particles with certain scalar properties ($p_t$, $E$, etc.), another collection of jets, a collection of tracks, and so on.
A typical physics analysis uses a large number of events but processes only a subset of the available properties.
Therefore, ROOT's TTree storage format support a \emph{columnar} physical data layout for nested sub records and collections~\cite{root96}.
Values of a single property of many events (e.g., $p_t$ for events 1 to 1000) are stored consecutively on disk.
Thus, only those parts that are required for an analysis need to be read.
Similar values are likely to be grouped together, which is beneficial for compression.

More than \SI{1}{\exa\byte} of data is stored in the TTree format.
For HEP use cases, the TTree I/O speed and storage efficiency has shown to be significantly better than many industry products~\cite{hepformats18}.
Furthermore, ROOT provides the unique feature of seamless C++ and Python integration where users do not need to write or generate a data schema.
Yet, the TTree implementation limits the optimal use of new storage systems and storage device classes, such as object stores and flash memory,
and it shows shortcomings when it comes to multi-threaded and GPU supported analysis tasks and fail-safe APIs.

In this contribution, we present the design and first benchmarks of the \emph{RNTuple} set of classes.
The RNTuple classes provide a new, experimental columnar event I/O system that is backwards-incompatible to TTree both on the file format level and on the API level.
Breaking backwards compatibility allows us to use contemporary nomenclature and
to design the ROOT event I/O from the ground up for next-generation devices and the increased data rates expected from HL-LHC.

\section{Design of the RNTuple I/O subsystem}
\label{sct:design}

This section describes key design choices of the RNTuple data format and of class design and the interfaces.

\subsection{Data layout}

\begin{figure}[h]
     \centering
     \resizebox{12cm}{!}{\begin{tikzpicture}
  \node[rectangle, fill=gray, minimum width=1.25cm] (header) {};
	\node[rectangle, fill=cyan, minimum width=1cm, anchor=west] (colId) at (header.east) {};
  \node[rectangle, fill=red, minimum width=1cm, anchor=west] (colP) at (colId.east) {};
  \node[rectangle, fill=orange, minimum width=1cm, anchor=west] (colE1) at (colP.east) {};
  \node[rectangle, fill=magenta!75, minimum width=1cm, anchor=west] (colPidOff1) at (colE1.east) {};
  \node[rectangle, fill=black!40!green, minimum width=1cm, anchor=west] (colPid1) at (colPidOff1.east) {};
  \node[rectangle, fill=black!40!green, minimum width=1cm, anchor=west] (colPid2) at (colPid1.east) {};
  \node[rectangle, fill=orange, minimum width=1cm, anchor=west] (colE2) at (colPid2.east) {};
  \node[rectangle, fill=magenta!75, minimum width=1cm, anchor=west] (colPidOff2) at (colE2.east) {};
  \node[rectangle, fill=black!40!green, minimum width=1cm, anchor=west] (colPid3) at (colPidOff2.east) {};
  \node[rectangle, minimum width=2cm, anchor=west] (etc) at (colPid3.east) {\dots};
  \node[rectangle, fill=cyan, minimum width=1cm, anchor=west] (colIdagain) at (etc.east) {};
  \node[rectangle, minimum width=2cm, anchor=west] (etc2) at (colIdagain.east) {\dots};
  \node[rectangle, fill=gray, minimum width=1.5cm, anchor=west] (footer) at (etc2.east) {};

  \draw[->, black!75, curve to, out=-45, in=225] ($(colP.south west) + (0.5ex, 0)$) to ($(colE1.south west) + (0.5ex, 0)$);
  \draw[->, black!75, curve to, out=-45, in=225] ($(colP.south west) + (0.5ex, 0)$) to ($(colPidOff1.south west) + (0.5ex, 0)$);
  \draw[->, black!75, curve to, out=10, in=170] ($(colP.north west) + (3.5ex, 0)$) to ($(colE2.north west) + (0.5ex, 0)$);
  \draw[->, black!75, curve to, out=10, in=170] ($(colP.north west) + (3.5ex, 0)$) to ($(colPidOff2.north west) + (0.5ex, 0)$);
  \draw[->, black!75, curve to, out=-45, in=225] ($(colPidOff1.south west) + (0.5ex, 0)$) to ($(colPid1.south west) + (0.5ex, 0)$);
  \draw[->, black!75, curve to, out=10, in=170] ($(colPidOff1.north west) + (2.5ex, 0)$) to ($(colPid2.north west) + (0.5ex, 0)$);
  \draw[->, black!75, curve to, out=-45, in=225] ($(colPidOff2.south west) + (0.5ex, 0)$) to ($(colPid3.south west) + (0.5ex, 0)$);

  \foreach \x in {1,...,6} {
    \draw[black!75] ($(colId.north west)+(\x ex, -1pt)$) -- ($(colId.south west)+(\x ex, 1pt)$);
  }
  \foreach \x in {1,...,6} {
    \draw[black!75] ($(colP.north west)+(\x ex, -1pt)$) -- ($(colP.south west)+(\x ex, 1pt)$);
  }
  \foreach \x in {1,...,6} {
    \draw[black!75] ($(colE1.north west)+(\x ex, -1pt)$) -- ($(colE1.south west)+(\x ex, 1pt)$);
  }
  \foreach \x in {1,...,6} {
    \draw[black!75] ($(colE2.north west)+(\x ex, -1pt)$) -- ($(colE2.south west)+(\x ex, 1pt)$);
  }
  \foreach \x in {1,...,6} {
    \draw[black!75] ($(colPidOff1.north west)+(\x ex, -1pt)$) -- ($(colPidOff1.south west)+(\x ex, 1pt)$);
  }
  \foreach \x in {1,...,6} {
    \draw[black!75] ($(colPidOff2.north west)+(\x ex, -1pt)$) -- ($(colPidOff2.south west)+(\x ex, 1pt)$);
  }
  \foreach \x in {1,...,6} {
    \draw[black!75] ($(colPid1.north west)+(\x ex, -1pt)$) -- ($(colPid1.south west)+(\x ex, 1pt)$);
  }
  \foreach \x in {1,...,6} {
    \draw[black!75] ($(colPid2.north west)+(\x ex, -1pt)$) -- ($(colPid2.south west)+(\x ex, 1pt)$);
  }
  \foreach \x in {1,...,6} {
    \draw[black!75] ($(colPid3.north west)+(\x ex, -1pt)$) -- ($(colPid3.south west)+(\x ex, 1pt)$);
  }
  \foreach \x in {1,...,6} {
    \draw[black!75] ($(colIdagain.north west)+(\x ex, -1pt)$) -- ($(colIdagain.south west)+(\x ex, 1pt)$);
  }

  \draw[thick, black!60, |-|] ($(header.west)-(-0.25ex, 1em)$) -- coordinate[midway] (coordHeader) ($(header.east)-(0.25ex, 1em)$);
  \node[minimum width=2cm, anchor=north] at ($(coordHeader.south)-(0, 0.5ex)$) {Header};

  \draw[thick, black!60, |-|] ($(footer.west)-(0ex, 1em)$) -- coordinate[midway] (coordFooter) ($(footer.east)-(0ex, 1em)$);
  \node[minimum width=2cm, anchor=north] at ($(coordFooter.south)-(0, 0.5ex)$) {Footer};

  \draw[thick, black!60, |-|] ($(colId.west)-(0, 1em)$) -- coordinate[midway] (coordPage) ($(colId.east)-(0, 1em)$);
  \node[minimum width=2cm, anchor=north] at ($(coordPage.south)-(0, 0.5ex)$) {Page};

  \draw[thick, black!60, |-|] ($(colId.west)-(0, 4em)$) -- coordinate[midway] (coordCluster) ($(etc.east)-(0, 4em)$);
  \node[minimum width=2cm, anchor=north] at ($(coordCluster.south)-(0, 0.5ex)$) {Cluster};

  \draw[thick, black!60, |-|] ($(header.west)+(0, 2em)$) -- coordinate[midway] (coordFile) ($(footer.east)+(0, 2em)$);
  \node[minimum width=2cm, anchor=south] at ($(coordFile.north)+(0, 0ex)$) {Dataset / File};

  \node[align=left, fill=black!10, rounded corners, font=\scriptsize, anchor=north west, xshift=0cm, yshift=-1em] at (colPidOff1.south west) {
    C++ collections become offset columns
  };


  \node[align=left, fill=black!10, rounded corners, font=\tt, anchor=north east, xshift=3cm, yshift=-3em] at (etc2.south east) {
    {\bf struct} Event \{\\
    \ \ \textcolor{cyan}{{\bf int} fId};\\
    \ \ \textcolor{red}{{\bf vector}$<$}Particle\textcolor{red}{$>$ fPtcls};\\
    \};\\
    {\bf struct} Particle \{\\
    \ \ \textcolor{orange}{{\bf float} fE};\\
    \ \ \textcolor{magenta!75}{{\bf vector}$<$}\textcolor{black!40!green}{int}\textcolor{magenta!75}{$>$ fIds};\\
    \};
  };

  \node[draw, fill=white, rounded corners, thick, font={\footnotesize}, text width=8cm, yshift=-2cm, anchor=north west] at (coordPage) {
    {\bf Approximate translation between TTree and RNTuple concepts:}\\[1em]
               \begin{tabular}{llcl}
                 & Basket     & $\approx$ & Page\\
                 & Leaf       & $\approx$ & Column\\
                 & Cluster    & $\approx$ & Cluster\\
               \end{tabular}
    };
\end{tikzpicture}

}
     \caption{Breakdown of the RNTuple data layout.
          Each scalar field of the event struct is stored in a separate column.}
     \label{fig:format}
\end{figure}
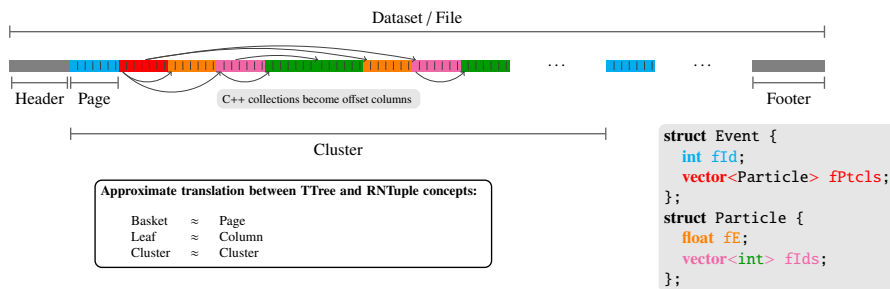

Compared to the TTree binary data layout, the RNTuple data layout is modestly modernized and borrows some ideas from Apache Arrow~\cite{sw-arrow} (see~Figure~\ref{fig:format}).
Data is stored in \emph{columns} of fundamental types supporting arbitrarily deeply nested collections (TTree drops the ``columnar-ness'' for deeply nested collections).%
\footnote{In contrast to TTree, RNTuple currently does not support row-wise storage.}
Columns are partitioned in compressed \emph{pages}, of typically a few tens of kilobytes in size.
Like in TTree, \emph{clusters} are a set of pages that contain all the data of a certain event range.
They are typically a few tens of megabytes in size and a natural unit of processing for a single thread or task.

A collection's representation contains an offset column whose elements indicate the start index within the columns that store the collection content; this allows for random-access of individual events.
The indexing is local to the cluster such that clusters can be written in parallel and freely concatenated to a larger data set.
This also allows for ``fast merging'', where several RNTuple files can be concatenated by only adjusting the header and footer.
In contrast to TTree, offset pages and value pages are always separated, which should improve the compression ratio (to be confirmed).
Integers and floating point numbers in columns are stored in little-endian format (TTree: big-endian) in order to allow for memory mapping of pages on most contemporary architectures.
Boolean values, such as trigger bits, are stored as bitmaps (TTree: byte arrays), which improves the compression.

The RNTuple meta-data are stored in a header and a footer.
The header contains the schema of the RNTuple; the footer contains the locations of the pages.
At a later point, we will extend the meta-data with a regularly written \emph{checkpoint footer} (e.\,g.~every \SI{100}{\mega\byte})
in order to allow for data recovery in case of an application crash during data taking.
We will also extend the meta-data with a user-accessible, namespace-scoped map of key-value pairs, such that the experiment data management systems can maintain relevant information
(checksums, replica locations, etc.) together with the data.

The pages, header and footer do not necessarily need to be written consecutively in a single file.
The container for pages, header and footer can be a ROOT file where data is interleaved with other objects such as histograms.
The container can also be an RNTuple bare file or an object store.
It is also conceivable to store header and footer in a different file than the pages to avoid backward seeks.

\subsection{Class design}

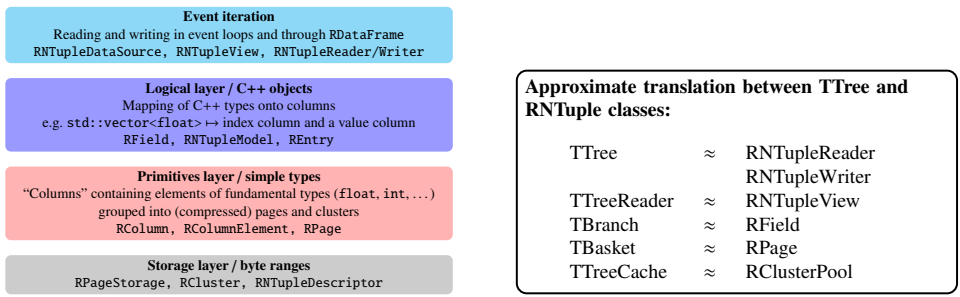
\begin{figure}[h]
     \centering
     \resizebox{6cm}{!}{
          \begin{tikzpicture}
            \node[rectangle, rounded corners, fill=black!20, minimum width=11cm, align=center] (storage) {\bf{Storage layer / byte ranges}\\
              \tt RPageStorage, RCluster, RNTupleDescriptor};
            \node[rectangle, rounded corners, fill=red!30, minimum width=11cm, anchor=south, yshift=1em, align=center] (phys) at (storage.north)
              {{\bf Primitives layer / simple types}\\``Columns'' containing elements of fundamental types (\texttt{float}, \texttt{int}, \dots)\\grouped into (compressed) pages and clusters\\
              \tt RColumn, RColumnElement, RPage};
            \node[rectangle, rounded corners, fill=blue!40, minimum width=11cm, anchor=south, yshift=1em, align=center] (logical) at (phys.north)
              {{\bf Logical layer / C++ objects}\\Mapping of C++ types onto columns\\
              e.g.~\texttt{std::vector<float>} $\mapsto$ index column and a value column\\
              \tt RField, RNTupleModel, REntry};
            \node[rectangle, rounded corners, fill=cyan!40, minimum width=11cm, anchor=south, yshift=1em, align=center] (iteration) at (logical.north)
              {\bf{Event iteration} \\
              Reading and writing in event loops and through {\tt RDataFrame}\\
              \tt RNTupleDataSource, RNTupleView, RNTupleReader/Writer};
          \end{tikzpicture}
     }\quad\quad
     \resizebox{6cm}{!}{
          \begin{tikzpicture}
               \node[xshift=2cm, draw, opacity=1, anchor=west, align=left, fill=white, rounded corners, thick, font={\footnotesize}, text width=6cm]
                 at (current page.west) {
               {\bf Approximate translation between TTree and RNTuple classes:}\\[1em]
               \begin{tabular}{llcl}
                 & TTree       & $\approx$ & RNTupleReader\\
                 &             &           & RNTupleWriter\\
                 & TTreeReader & $\approx$ & RNTupleView\\
                 & TBranch     & $\approx$ & RField\\
                 & TBasket     & $\approx$ & RPage\\
                 & TTreeCache  & $\approx$ & RClusterPool\\
               \end{tabular}
               };
          \end{tikzpicture}
     }
     \caption{Layers and key class of RNTuple and their approximate counterparts in TTree.}
     \label{fig:design}
\end{figure}

The RNTuple class design comprises four layers (see~Figure~\ref{fig:design}).
The RNTuple classes make use of templates, such that for simple types (e.g., vectors of floats) that are known at compile time,
the compiler can inline a fast path from the highest to the lowest layer without additional value copies or virtual calls.

The \emph{event iteration layer} provides the user-facing interfaces to read and write events, either through RDataFrame~\cite{dataframe18} or as hand-written event loops.
The user interface is presented in more detail in Section~\ref{sct:interface}.

The \emph{logical layer} splits C++ objects into columns of fundamental types.
Its central class is the \emph{RField} that provides a C++ template specialization for reading and writing of an I/O supported type.
Currently there is support for
boolean, integer and floating point, \texttt{std::vector} and \texttt{std::array} containers, \texttt{std::string}, \texttt{std::variant}, and user-defined classes with a ROOT dictionary.
In the future, we will provide support for addtional types (e.g., \texttt{std::map}, \texttt{std::chrono}) and possibly for intra-event object references as a limited form of pointers.
While RNTuple limits I/O support to an explicit subset of C++ types, those types are fully composable (e.g., a user-defined class containing a vector of arrays of another user-defined class).

The \emph{primitives layer} governs the pool of uncompressed and deserialized pages in memory and the representation of fundamental types on disk.
For most fundamental types, the memory layout equals the RNTuple on-disk layout.
In some circumstances, pages need to be \emph{packed} and \emph{unpacked}, for instance in order to store booleans as bitmaps or in order to store floating point values with
reduced precision.

The \emph{storage layer} provides access to the byte ranges containing a page on a physical or virtual device.
The storage layer manages compression and reads and writes to and from the I/O device.
It also allocates memory for pages in order to allow for direct page mappings.
Currently there is support for a storage layer that uses a ROOT file as an RNTuple data container and a storage layer that uses a bare file for comparison and testing.
We plan to add another implementation that uses an object store.
We will also add virtual storage layers that combine RNTuple data sets similar to TTree's \emph{chains} and \emph{friend trees}.

An RNTuple \emph{cluster pool} provides I/O scheduling capabilities.
The cluster pool spawns an I/O thread that asynchronously preloads upcoming pages of active columns.
The cluster pool can linearize, merge and split requests to optimize the read pattern for the storage device at hand (e.\,g.~spinning disk, flash memory, remote server).

\section{RNTuple user interfaces}
\label{sct:interface}

The RNTuple user-facing API is supposed to be easy to use correctly as to minimize the likelihood of application crashes and wrong results.
To this end, RNTuple provides an RDataFrame data source so that RDataFrame analyses code can be use unmodified with RNTuple data.

The RNTuple interface for implementing hand-written event loops uses modern standard techniques,
including smart pointers, event traversal by C++ iterators and compile-time safety through templated interfaces~(see~Figure~\ref{fig:interface}).
For the type-unsafe interface, a runtime check verifies that the the on-disk type and the in-memory type of fields match.

\begin{figure}[h]
     \begin{center}
     \begin{minipage}[t]{6cm}
          \scriptsize
          \begin{minted}[bgcolor=gray!30]{c++}
auto ntpl =
   RNTupleReader::Open("Events", "f.root");
auto viewPt = ntpl->GetView<float>("pt");

for (auto i : ntpl->GetEntryRange()) {
   hist.Fill(viewPt(i));
}
          \end{minted}
     \end{minipage}
     \quad
     \begin{minipage}[t]{6cm}
          \scriptsize
          \begin{minted}[bgcolor=gray!30]{c++}
auto model = RNTupleModel::Create();
auto fldPt = model->MakeField<float>("pt");
// Note: there is also a void* based,
//       runtime type-safe API

auto ntpl = RNTupleReader::Open(
   std::move(model), "Events", "f.root");

for (auto entryId : *ntpl) {
   ntuple->LoadEntry(entryId);
   hist.Fill(*fldPt);
}
          \end{minted}
     \end{minipage}
     \end{center}
     \caption{RNTuple interface sketch for reading data.
     Left-hand side: zero-copy interface where the memory pointed to by \emph{views} is managed by RNTuple.
     Right-hand side: interface that copies values into generated or user-provided memory locations.}
     \label{fig:interface}
\end{figure}

The RNTuple classes are thread-friendly, i.\,e.~multiple threads can safely use their own copy of RNTuple classes to read the same data concurrently.
In the future, we envision support for multi-threaded writing (one cluster per thread or task)
as well as support for multiple threads reading concurrently from the same range of clusters of an RNTuple.
In single-threaded analyses, available idle cores should be used for decompression.
We believe that these changes will require very little changes to the user-facing API.

Error handling, for instance in case of device faults or malformed input data, is an important aspect of I/O interfaces.
While it is often difficult to recover gracefully from I/O errors, the I/O layer should reliably detect errors and produce an error report as close as possible to the root cause.
To this end, RNTuple throws C++ exceptions for I/O errors.

At a later point, we intend to add a limited C API for RNTuple in order to facilitate ROOT data being transferred to 3rd party consumers, such as numpy arrays or machine learning toolkits.
To this end, most of RNTuple is implemented not to depend on core ROOT classes, such that a minimal, stand-alone RNTuple I/O library can be built.
The functionality of this library will initially be limited to reading simple numerical type fields and vectors thereof.

\section{Performance evaluation}

In this section, we analyze the RNTuple performance in terms of read throughput and file size for typical, single threaded analysis tasks.
We use three sample analyses for the benchmarks (see~Table~\ref{tab:benchmarks}).
Each analysis requires a subset of the available event properties, uses some properties to filter events, and calculates an invariant mass from the selected events.
The analyses were implemented using both TTree and RNTuple, each variant optimized for best performance with hand-written event loops\footnote{For
the implementation, see \url{https://github.com/jblomer/iotools/tree/ntuple-chep-2019}}.
Basket/Page sizes and cluster sizes are comparable between TTree and RNTuple files.
The ``LHCb'' sample is derived from an LHCb Open Data course~\cite{lhcbopendata16}.
The ``H1'' sample is derived from the ROOT ``H1 analysis'' tutorial with the original data cloned ten times.
The ``CMS'' sample is derived from the ROOT ``dimuon'' tutorial using the 2019 nanoAOD format~\cite{nanoaod19} with simulated data.
Two dedicated physical nodes, ``machine 1'' and ``machine 2'' are used for running the benchmarks~(see~Table~\ref{tab:hardware}).
Both machines run CentOS 7 and have ROOT\footnote{ROOT branch \url{https://github.com/jblomer/root/tree/ntuple-chep-2019}} compiled with gcc 7.3.
A third dedicated node runs XRootD in version 4.10 and is configured to hold the data on a RAM disk.

\begin{table}
     \centering
     \caption{Sample analyses for performance evaluation.
     Main differences are the data model (flat or with collections) and the number of required branches (dense or sparse reading).}
     \label{tab:benchmarks}
     \begin{tabular}{lll}
     \hline
     LHCb run 1 open data B2HHH              & H1 micro dst [$\times$10]                  & CMS nanoAOD June 2019\\\hline
     18/26 branches (>\SI{75}{\percent})     & 16/152 branches ($\sim$\SI{10}{\percent})  & 6/1479 branches (<\SI{1}{\percent}) \\
     fully flat data model                   & event sub collections                      & event sub collections \\
     8.5 million events                      & 2.8 million events                         & 1.6 million events\\
     24\,k selected events                   & 75\,k selected events                      & 141\,k selected events\\\hline
     \end{tabular}
\end{table}

\begin{table}
     \centering
     \caption{Overview of the benchmarking hardware.}
     \label{tab:hardware}
     \begin{tabular}{lll}
          \hline
          Hardware                  & Machine 1                                    & Machine 2 \\\hline
          CPU                       & Xeon Platinum 8260 @ \SI{2.4}{\giga\hertz}   & Xeon E5-2630v3 @ \SI{2.4}{\giga\hertz} \\
          Memory                    & DDR4 RDIMM \SI{2933}{\mega\hertz}            & DDR4 RDIMM \SI{2133}{\mega\hertz} \\
          Optane (NVRAM)            & Optane DC \SI{2666}{\mega\hertz} (ext4/DAX)  & --- \\
          SSD (flash)               & Intel DC P4510, PCIe 3.1 $\times$4           & --- \\
          HDD (spinning)            & ---                                          & 2$\times$ SAS 7200\,RPM (RAID1)\\
          Network                   & ---   & 1\,GbE \\\hline
     \end{tabular}
\end{table}

\subsection{Storage efficiency}

\begin{figure}[h]
     \centering
     \includegraphics[width=4.25cm]{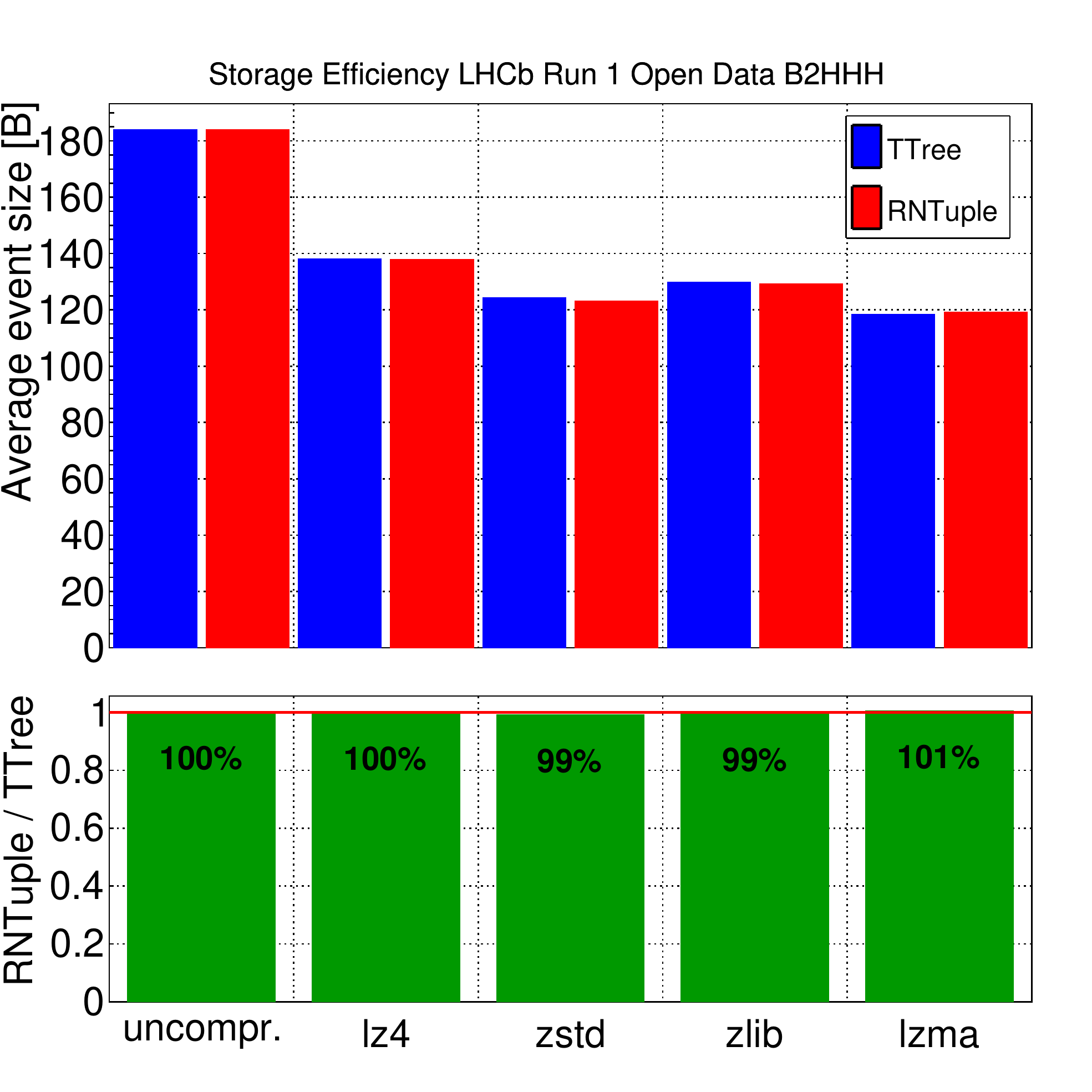}
     \includegraphics[width=4.25cm]{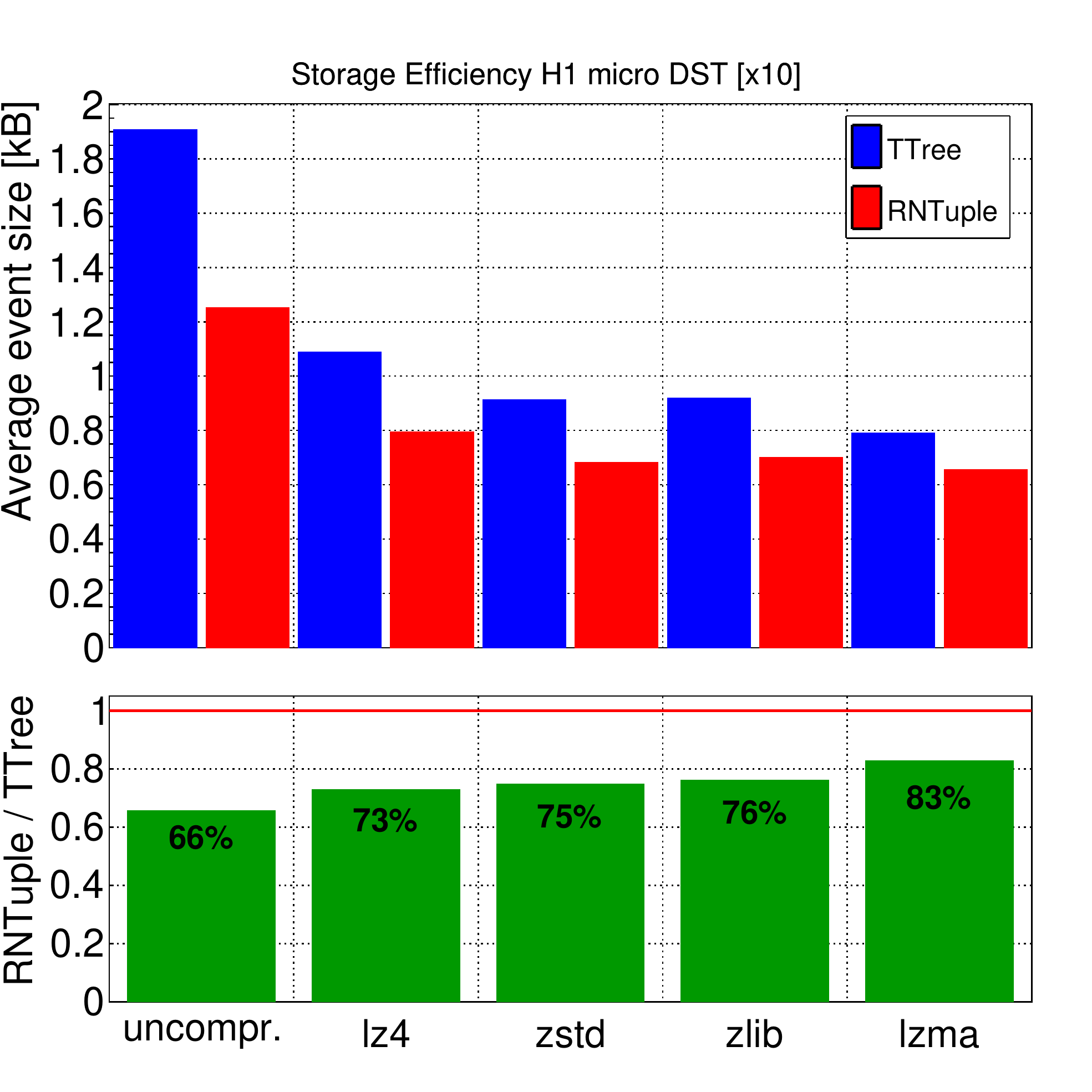}
     \includegraphics[width=4.25cm]{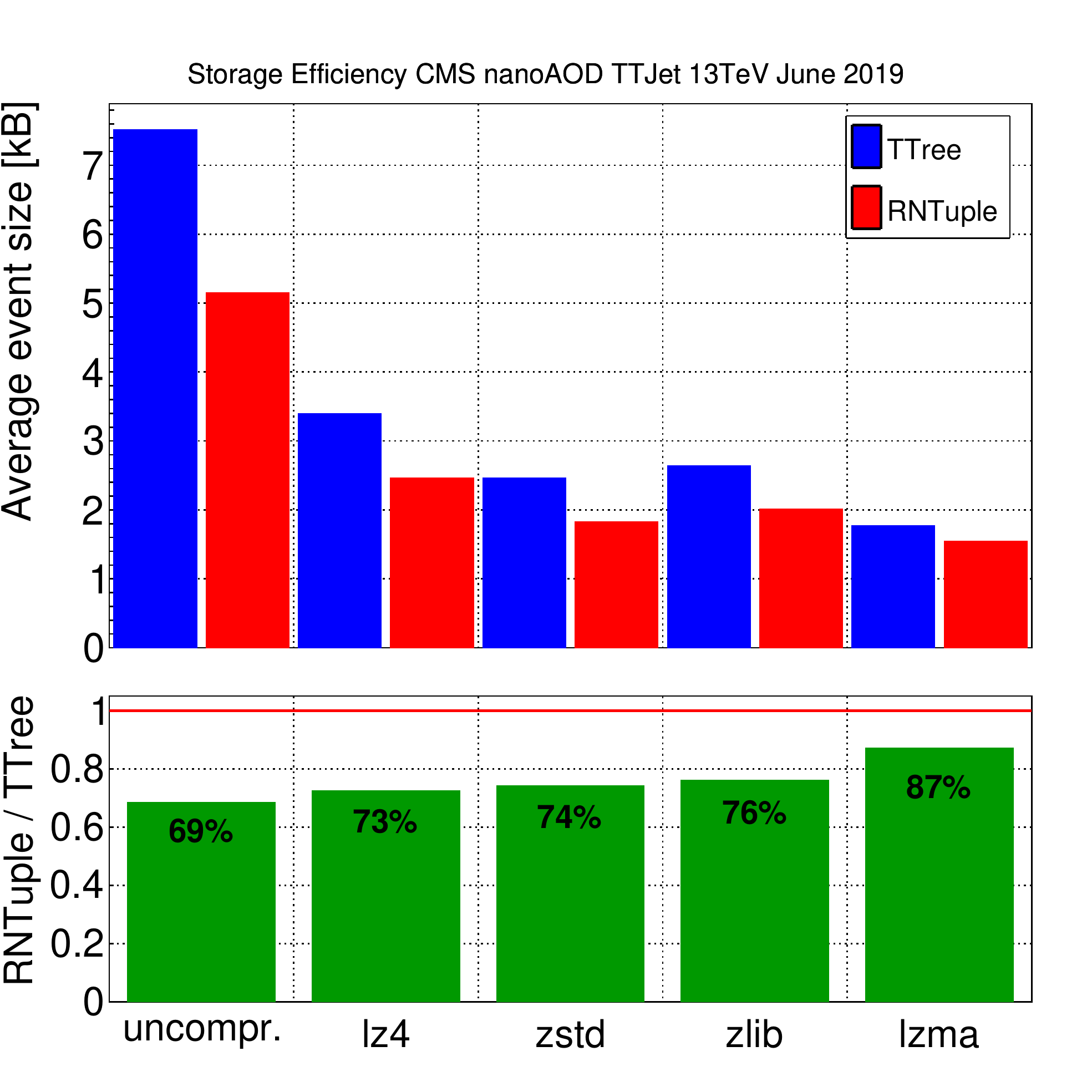}
     \caption{File size comparison of the sample analysis input data in TTree and RNTuple format with different compression algorithms.}
     \label{fig:size}
\end{figure}

Figure~\ref{fig:size} shows the file format efficiency for the input data of the sample analysis.
As expected, the TTree and RNTuple efficiency is very similar on the ``LHCb'' flat data model.
For ``H1'' and ``CMS'', RNTuple shows significantly better efficiency due to the more efficient storage of collections and boolean values.
(Approximately half of the difference in file size could be eliminated by using TTree's experimental \texttt{kGenerateOffsetMap} I/O flag.)
Space savings of RNTuple remain even after compression.

\subsection{Read performance}

\begin{figure}[h]
     \centering
     \includegraphics[width=4.25cm]{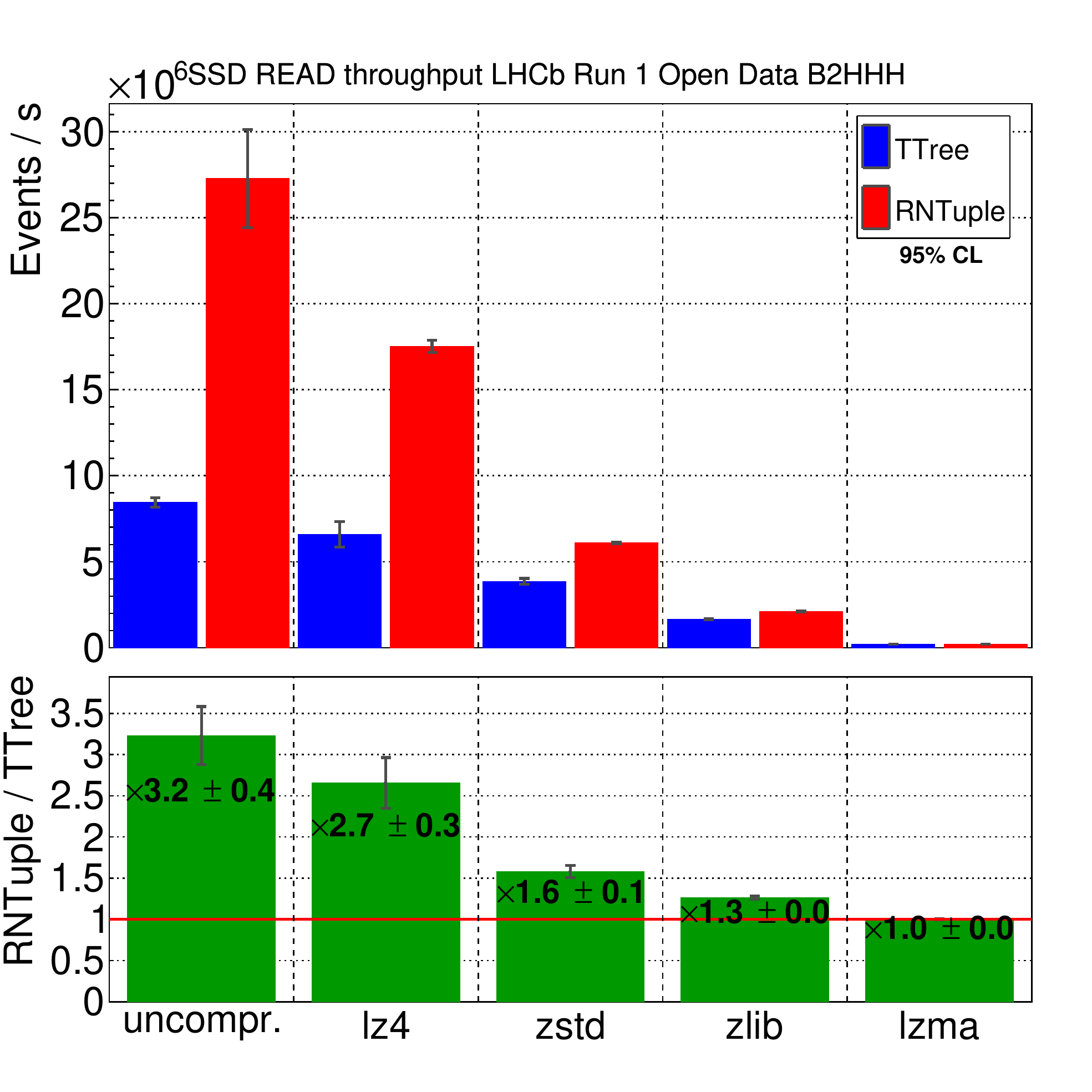}
     \includegraphics[width=4.25cm]{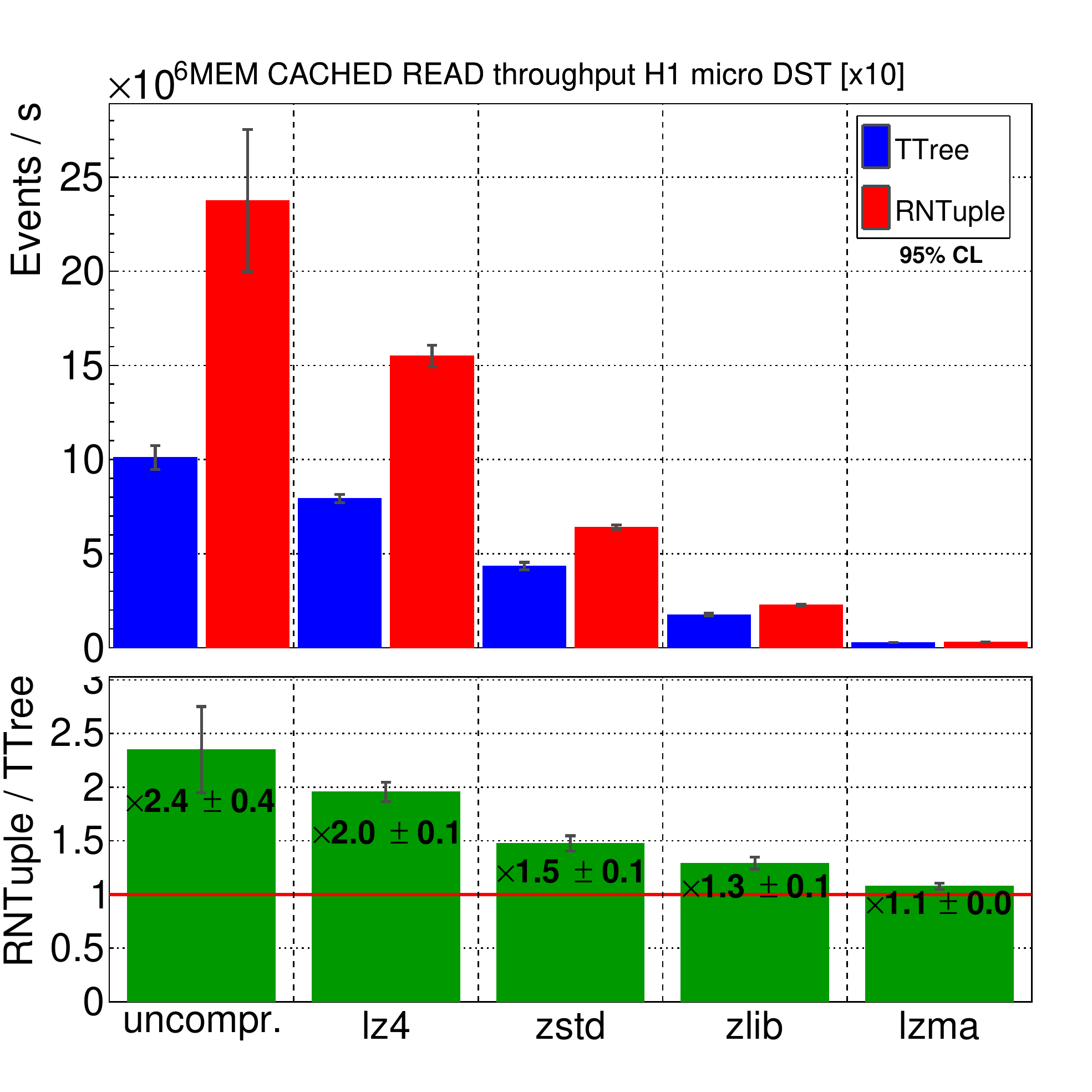}
     \includegraphics[width=4.25cm]{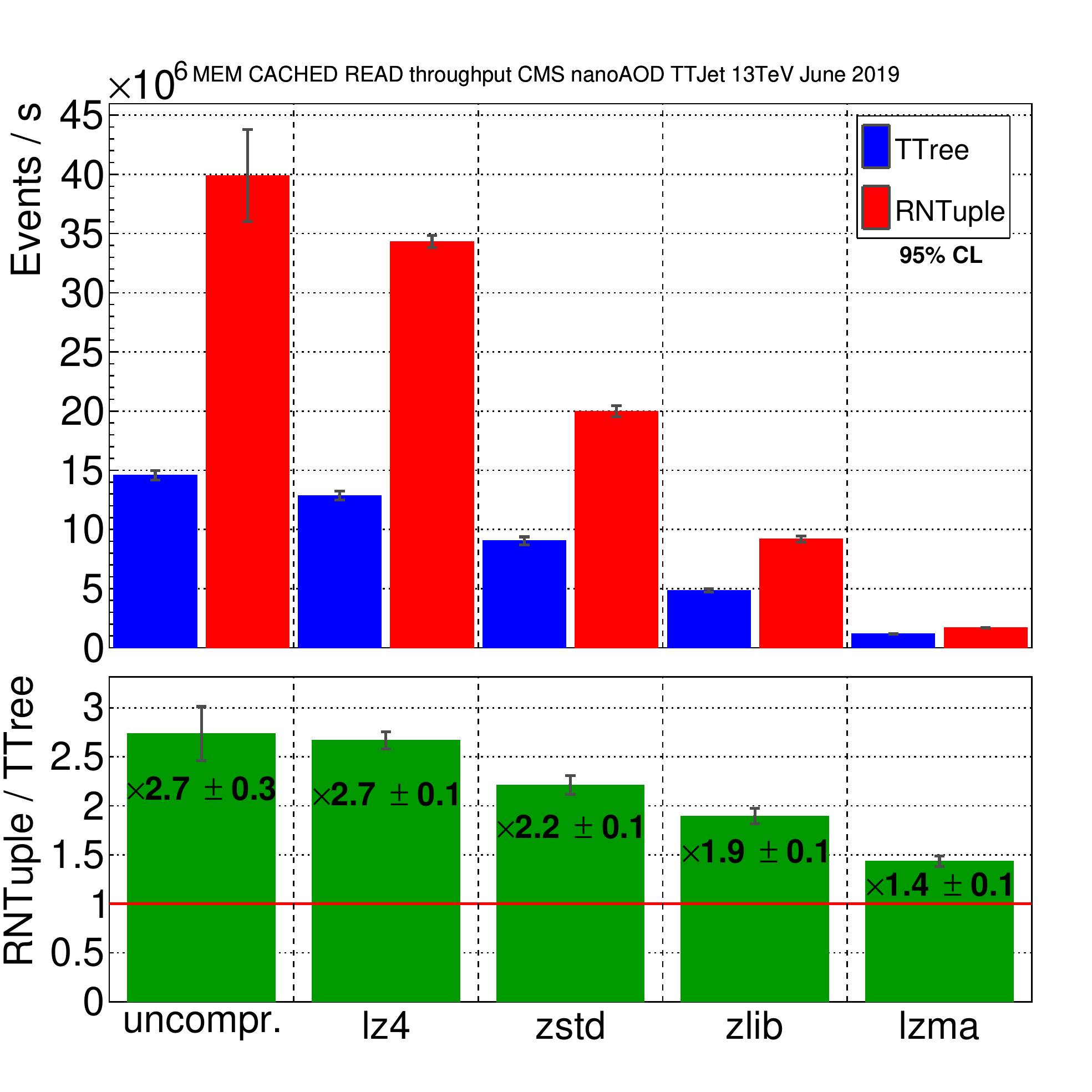}
     \\
     \includegraphics[width=4.25cm]{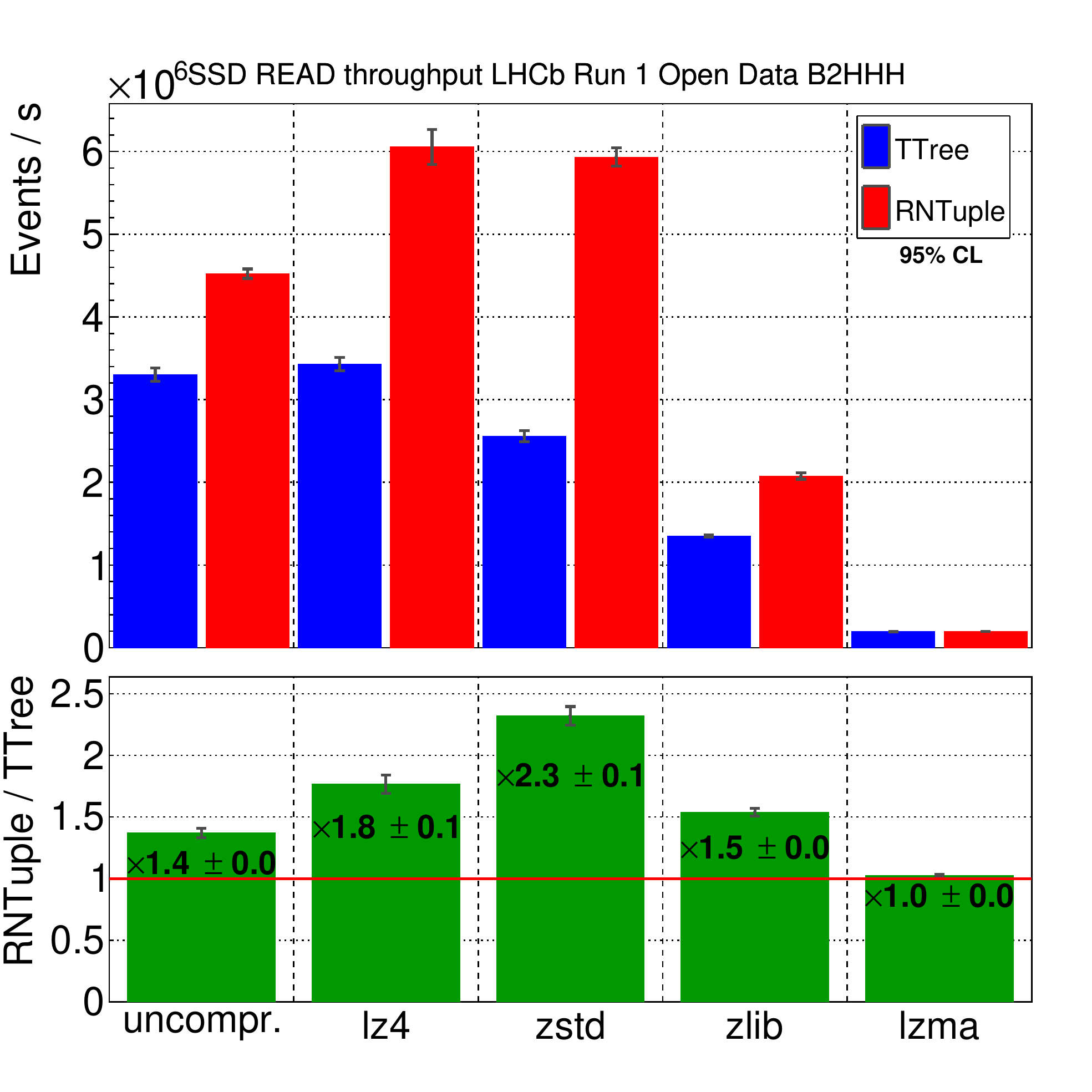}
     \includegraphics[width=4.25cm]{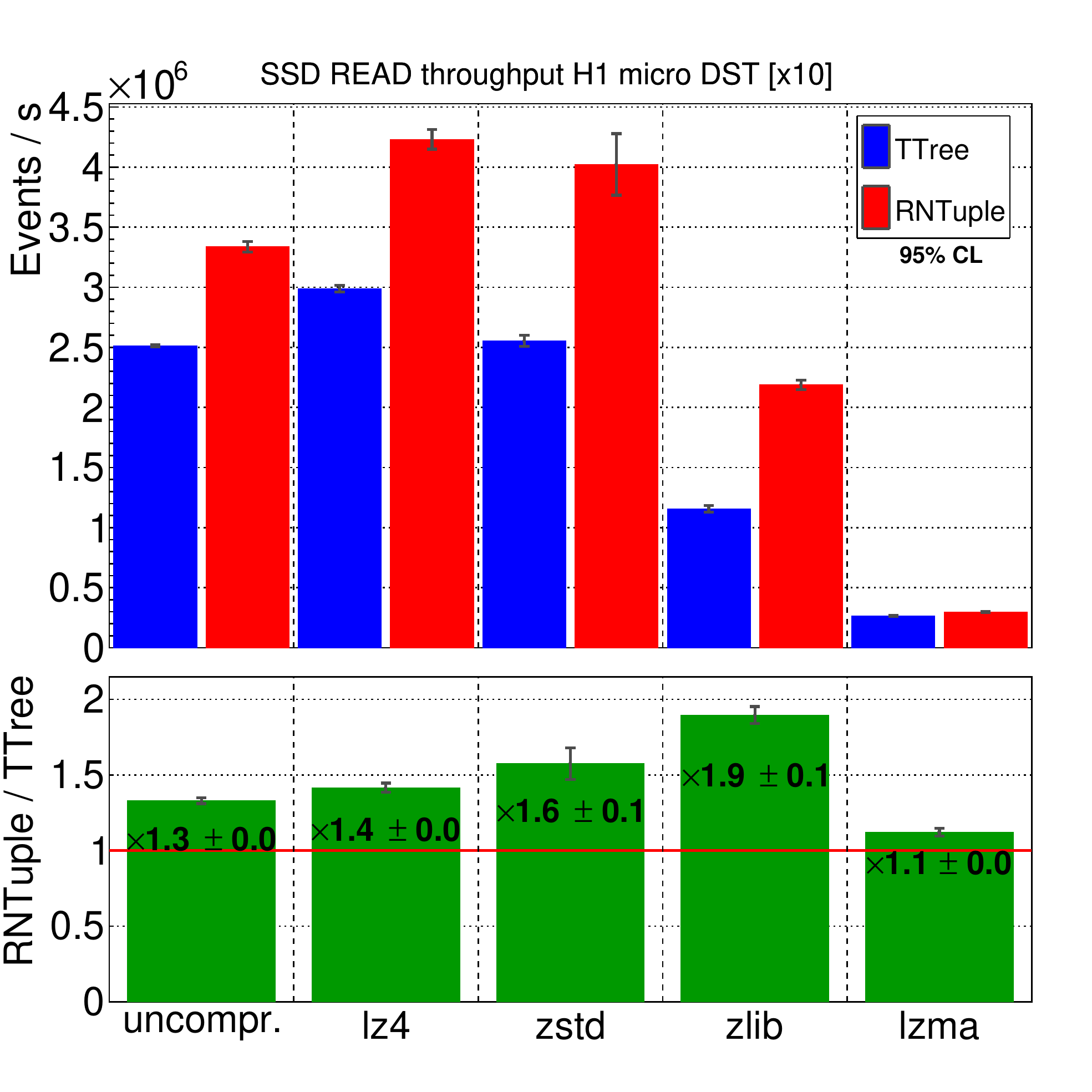}
     \includegraphics[width=4.25cm]{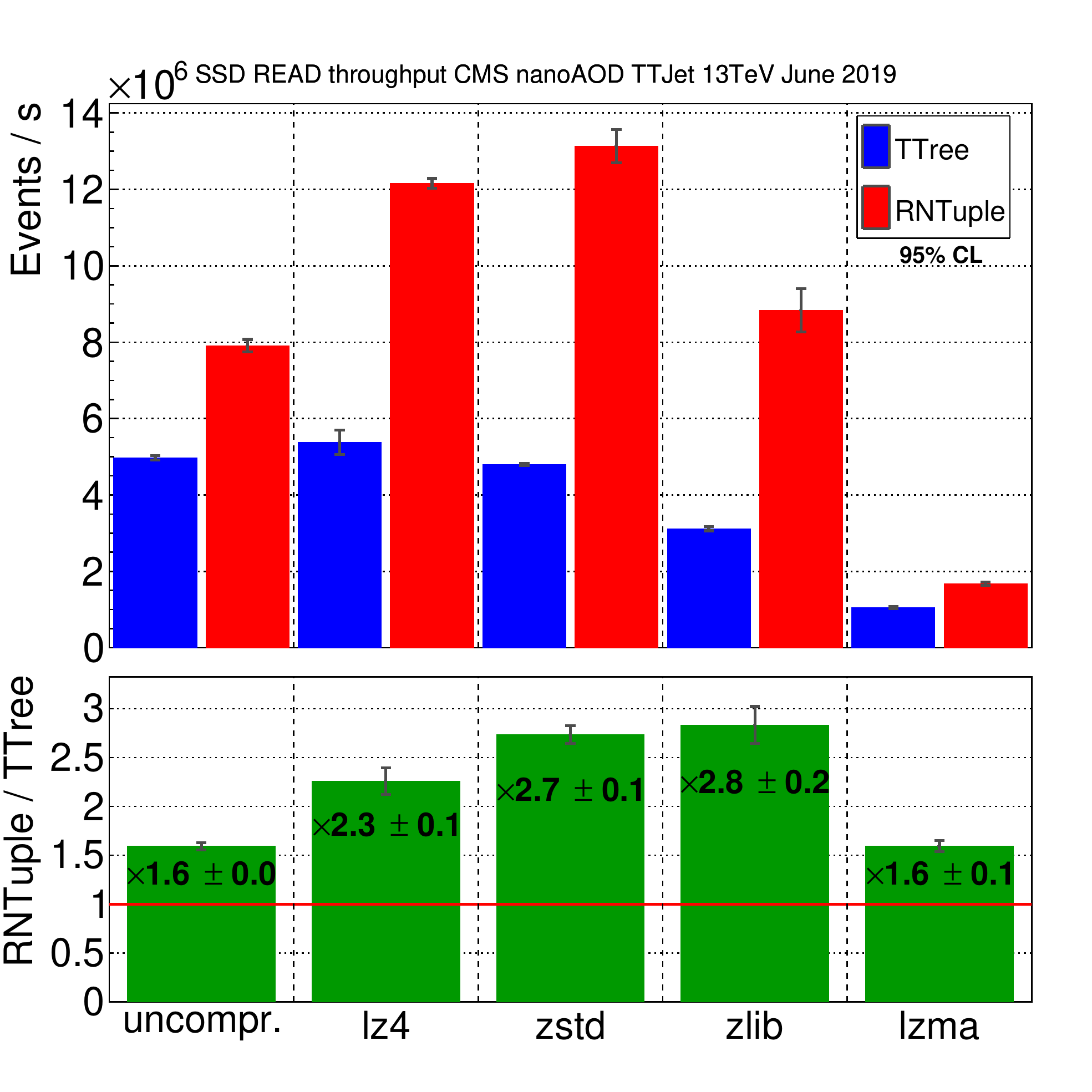}
     \caption{Read throughput in events per second on machine 1.
          Upper half shows the results for warm file system buffers.
          Lower half shows the results for reading from SSD with a cold cache.
          See Figure~\ref{fig:streams} for further improvements for SSDs.}
     \label{fig:read}
\end{figure}

Figure~\ref{fig:read} shows the event throughput for running the sample analyses.
When reading from warm file system buffers, the performance is dominated by deserialization and decompression.
The data deserialization in RNTuple is significantly faster compared to TTree.
With stronger compression algorithms, the performance is more dominated by decompression than by deserialization.
Still, even for LZMA compressed data reading RNTuple data is faster for the H1 and CMS samples.

When reading with cold file system buffers, as shown in the lower half of Figure~\ref{fig:read},
the performance depends not only on the deserialization and decompression speed but also on the I/O throughput of the device.
The additional CPU time spent on strong compression can be more than compensated by a smaller transfer volume.
For RNTuple, there is a sweet spot for the recent zstd compression algorithm,
in particular if taking into account the smaller file size as compared to zlib and lz4.

\begin{figure}[h]
     \centering
     \includegraphics[width=12cm]{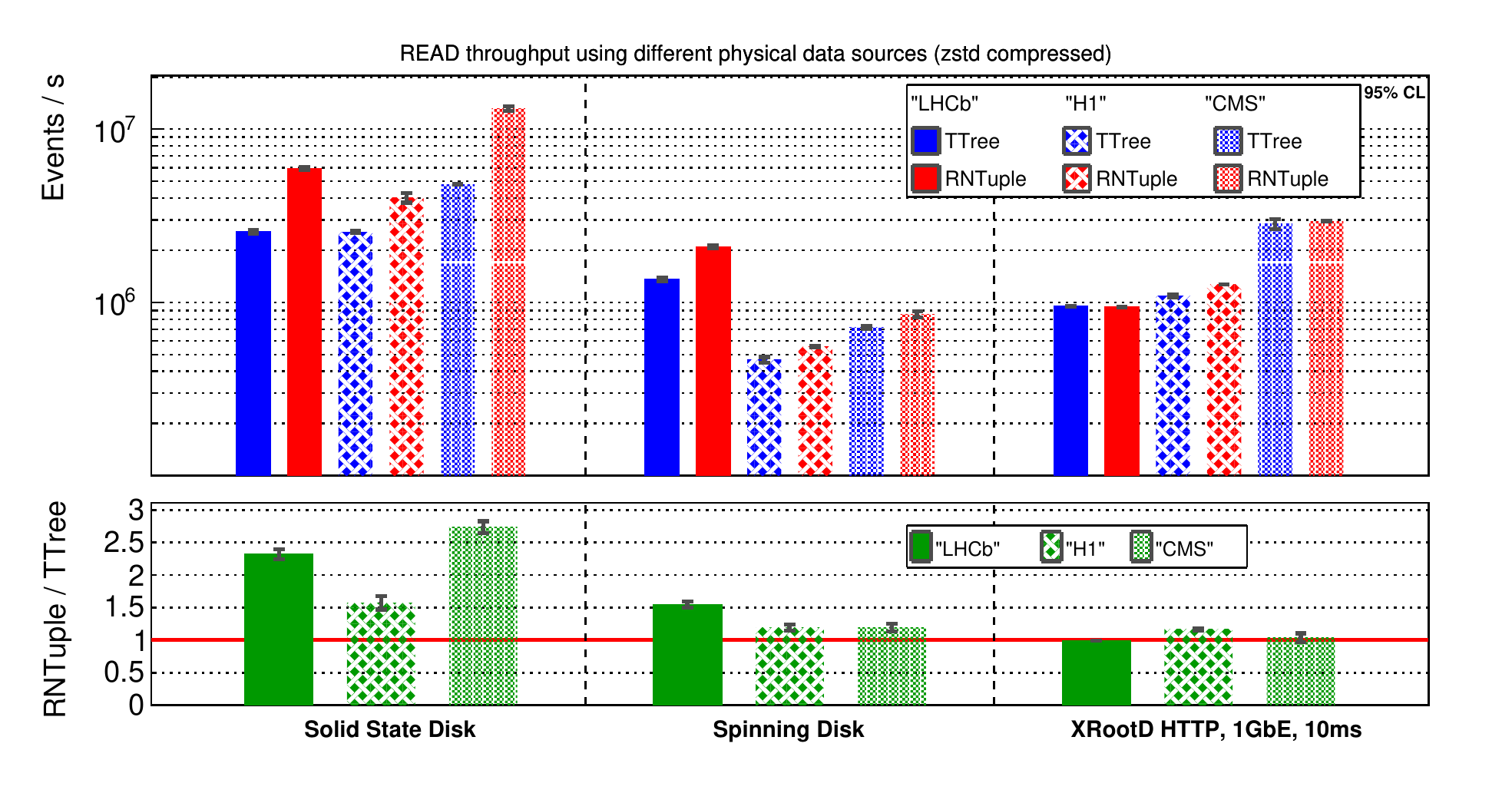}
     \caption{Read speed with different bandwidth and latency profiles.
          SSD benchmarks from machine 1, HDD and HTTP benchmarks from machine 2 connected to a dedicated, third XRootD server.
          Note that the SSD results on the left hand side are identical to the SSD/zstd results in Figure~\ref{fig:read}.}
     \label{fig:media}
\end{figure}

Figure~\ref{fig:media} compares cold cache read performance for different, frequently used physical data sources.
For the slow devices HDD and 10\,GbE, the performance is dominated by the I/O scheduler, i.\,e.~by TTreeCache resp.~RClusterPool.
The I/O scheduler linearizes requests, merges nearby requests, and issues vector reads in order to minimize the overall number of requests sent to a device and the total transfer volume.
In these benchmarks the RNTuple's I/O scheduler shows a performance at least as good as the TTreeCache.

\subsection{SSD optimizations}

\begin{figure}[h]
     \centering
     \includegraphics[width=12cm]{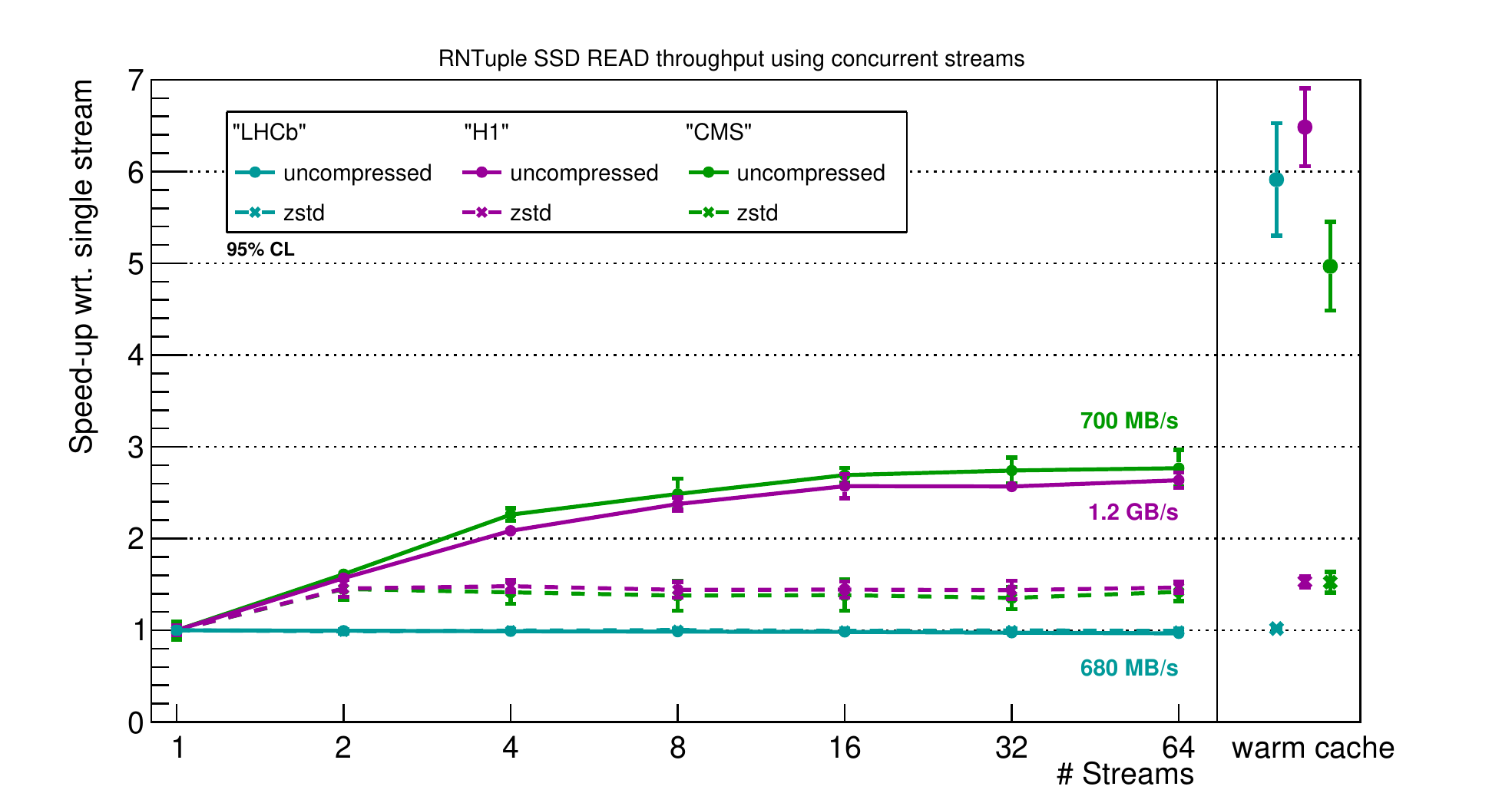}
     \caption{Full exploitation of SSDs by concurrent streams on machine 1.
          For comparison, the right hand side repeats the warm cache results from Figure~\ref{fig:read}.
          Single stream performance is identical to the SSD/zstd results in Figure~\ref{fig:read}.}
     \label{fig:streams}
\end{figure}

In contrast to spinning disks, SSDs are inherently parallel devices that benefit from a large queue depth so that they can read from multiple flash cells concurrently.
Figure~\ref{fig:streams} shows the effect of reading with multiple concurrent streams.
To this end, we extend the RNTuple I/O scheduler to read with multiple threads (1 stream/thread).
Where the read performance is limited by I/O and not by decompression and deserialization,
increasing the number of streams can yield another speed improvement of around a factor of 2.5.
The gains max out at around 16 streams.
The lower gains for uncompressed LHCb and CMS samples are due to a limitation in the current RNTuple implementation that only preloads a single cluster.
It therefore does not provide enough concurrent requests to fill the parallel streams.
With implementation of multi-cluster read-ahead, this limitation is going to be removed.
An interesting topic of future work is investigating automatic ways of the I/O scheduler to adjust to the underlying physical hardware.

\subsection{Optane DC NV-RAM evaluation}

\begin{figure}[h]
     \centering
     \includegraphics[width=12cm]{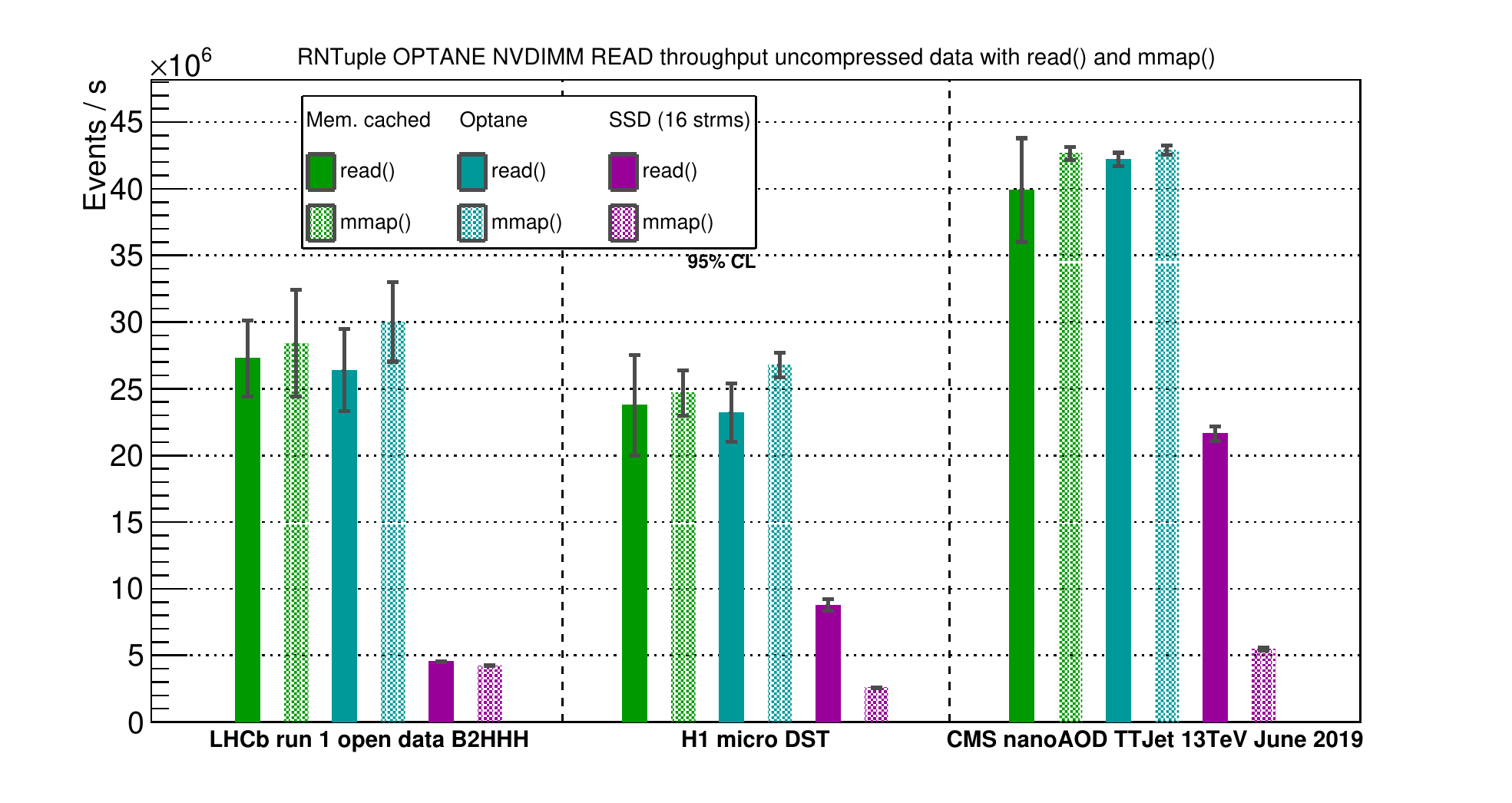}
     \caption{Read performance using Optane DC NV-RAM in ``App Direct'' mode on machine 1.
          The NV-RAM block device is formatted with ext4 with DAX optimization.
          Uncompressed input data is used in order to allow for comparsion between POSIX \texttt{read()} and \texttt{mmap()}.
          Note that the SSD \texttt{read()} results are identical to the 16 streams result in Figure~\ref{fig:streams}.}
     \label{fig:mmap}
\end{figure}

Figure~\ref{fig:mmap} shows the performance when reading RNTuple data from Optane DC NV-RAMs.
The performance characteristics of NV-RAMs are in-between RAM and SSDs (here, we are not exploiting the non-volatility).
In the future, they might become a more widespread additional cache layer or installed as a dedicated performance storage tier, e.\,g.~in analysis facilities.

The results show no significant difference between reading from warm file system caches and reading from NV-RAM.
As we also do not reach the peak throughput of the NV-RAM modules, the results suggest a bottleneck in the I/O deserialization or plotting part of the analysis run.
Further optimizations of the RNTuple I/O path are subject of future work.

Due to the fact that the RNTuple on-disk layout matches the in-memory layout, we can compare reading data explicitly with POSIX \texttt{read()} and implicitly by memory mapping.
On (byte-addressable) NV-RAM and warm file system buffers, both mechanisms yield comparable results.
When reading sparsely from SSDs, the RNTuple I/O scheduler optimizations bring a significant performance gain.
Further investigation reveals that the I/O scheduling that underpins the memory mapping in Linux issues an order of magnitude more requests to the device than the RNTuple scheduler.

\section{Conclusion}
\label{sct:final}

In this contribution we presented the design and a first performance evaluation of RNTuple, ROOT's new experimental event I/O system.
The RNTuple I/O system is a backwards-incompatible redesign of TTree, based on the many years of experience of the TTree development.
It is from the ground up designed to work well in concurrent environments
and to optimally support modern storage hardware and systems, such as SSDs, NV-RAM, and object stores.

Our benchmarks suggest that compared to TTree RNTuple can yield read speed improvements between a factor of 1.5 to 5 in realistic analysis scenarios,
while at the same time reducing data sizes by \SIrange{10}{20}{\percent}.
We will gradually move the RNTuple code from a prototype to a ROOT production component.
The RNTuple classes are already available in the \texttt{ROOT::Experimental::RNTuple} namespace if ROOT is compiled with the \texttt{root7} cmake option.
Tutorials are available to demonstrate the RNTuple functionality.
We consider these developments and the associated future R\&D topics essential building blocks for coping with data rates at the HL-LHC.

\section*{Acknowledgements}
\label{sct:ack}

We would like to thank Fons Rademakers and Luca Atzori from CERN openlab for giving us access to NV-RAM devices.
We would like to thank Dirk Düllmann and Michal Simon from CERN IT for providing us an XRootD test node.
We would like to thank Oksana Shadura, Brian Bockelman, and Jim Pivarski for many fruitful discussions and suggestions.



\bibliography{references.bib}

\end{document}